\documentclass{article}

     \PassOptionsToPackage{numbers, compress}{natbib}

     \usepackage[preprint]{neurips_2019}

\usepackage[utf8]{inputenc} 
\usepackage[T1]{fontenc}    
\usepackage{hyperref}       
\usepackage{url}            
\usepackage{booktabs}       
\usepackage{amsfonts}       
\usepackage{nicefrac}       
\usepackage{microtype}      
\usepackage[pdftex]{graphicx} 
\usepackage{amsmath}
\DeclareMathOperator*{\argmax}{argmax}
\usepackage[ruled]{algorithm2e}
\usepackage{float}
\title{Simulator-based training of generative models for the inverse design of metasurfaces}

\author{%
  Jiaqi Jiang \\
  Stanford University\\
  \texttt{jiangjq@stanford.edu} \\
  \And
  Jonathan A. Fan \\
  Stanford University \\
  \texttt{jonfan@stanford.edu} \\
}

\begin{document}

\maketitle

\begin{abstract}
Metasurfaces are subwavelength-structured artificial media that can shape and localize electromagnetic waves in unique ways. The inverse design of these devices is a non-convex optimization problem in a high dimensional space, making global optimization a major challenge. We present a new type of population-based global optimization algorithm for metasurfaces that is enabled by the training of a generative neural network.  The loss function used for backpropagation depends on the generated pattern layouts, their efficiencies, and efficiency gradients, which are calculated by the adjoint variables method using forward and adjoint electromagnetic simulations. We observe that the distribution of devices generated by the network continuously shifts towards high performance design space regions over the course of optimization. Upon training completion, the best generated devices have efficiencies comparable to or exceeding the best devices designed using standard topology optimization.  Our proposed global optimization algorithm can generally apply to other gradient-based optimization problems in optics, mechanics and electronics.
\end{abstract}

\section{Introduction}

Photonic technologies serve to manipulate, guide, and filter electromagnetic waves propagating in free space and in waveguides.  Due to the strong dependence of electromagnetic function on geometry, much emphasis in the field has been placed on identifying geometric designs for these devices given a desired optical response.  The vast majority of existing design concepts utilize relatively simple shapes that can be described using physically intuition.  For  example, silicon photonic devices typically utilize adiabatic tapers and ring resonators to route and filter guided waves \cite{Jalali2006}, and metasurfaces, which are diffractive optical components used for wavefront engineering, typically utilize arrays of nanowaveguides or nanoresonators comprising simple shapes \cite{Genevet2017}.  While these design concepts work well for certain applications, they possess limitations, such as narrow bandwidths and sensitivity to temperature, which prevent the further advancement of these technologies.

To overcome these limitations, design methodologies based on optimization have been proposed.  Among the most successful of these concepts is gradient-based topology optimization, which uses the adjoint variables method to iteratively adjust the dielectric composition of the devices and improve device functionality \cite{Vuckovic2018, Sigmund2011, Sawyer2019, sigmund2013topology, sigmund1997design, lalau2013adjoint}.  This design method, based on gradient descent, has enabled the realization of high performance, robust \cite{Wang2019} devices with nonintuitive layouts, including new classes of on-chip photonic devices with ultrasmall footprints \cite{sigmund2004, Vuckovic2015}, non-linear photonic switches \cite{SFan2018}, and diffractive optical components that can deflect \cite{sell2017large, JianjiOptExp2017, SellAdvOpt2017, JianjiAnnalen2018, SellACS2018} and focus \cite{Loncar2018, peter2019metalens} electromagnetic waves with high efficiencies.  While gradient-based topology optimization has great potential, it is a local optimizer and depends strongly on the initial distribution of dielectric material making up the devices \cite{Jianji2017}.  The identification of a high performance device is therefore computationally expensive, as it requires the optimization of multiple random initial dielectric distributions and selecting the best device.

We present a detailed mathematical discussion of a new global optimization concept based on Global Topology Optimization Networks (GLOnets) \cite{jiang2019global}, which combine adjoint variables electromagnetic calculations with the training of a generative neural network to realize high performance photonic structures.  Unlike gradient-based topology optimization, which optimizes one device at a time, our approach is population-based and optimizes a distribution of devices, thereby enabling a global search of the design space.  As a model system, we will apply our concept to design periodic metasurfaces, or metagratings, which selectively deflect a normally incident beam to the +1 diffraction order.  In our previous work \cite{jiang2019global}, we demonstrated that GLOnets conditioned on incident wavelength and deflection angle can generate ensembles of high efficiency metagratings.  In this manuscript, we examine the underlying mathematical theory behind GLOnets, specifically a derivation of the objective and loss functions, discussion of the training process, interpretation of hyperparameters, and calculations of baseline performance metrics for unconditional GLOnets.  We emphasize that our proposed concepts are general and apply broadly to design problems in photonics and other fields in the physical sciences in which the adjoint variables method applies.

\section{Related Machine Learning Work}

In recent years, deep learning has been investigated as a tool to facilitate the inverse design of photonic devices.  Many efforts have focused on using deep neural networks to learn the relationship between device geometry and optical response \cite{zhang2003, Figueroa2012}, leading to trained networks serving as surrogate models mimicking electromagnetic solvers.  These networks have been be used together with classical optimization methods, such as simulated annealing or particle swarm algorithms, to optimize a device \cite{cruz2013, Assuncao2010}.  Device geometries have also beens directly optimized from a trained network by using gradients from backpropagation \cite{peurifoy2018nanophotonic, inampudi2018neural, ma2018deep, liu2018training}.  These methods work well on simple device geometries described by a few parameters. However, the model accuracy decreases as the geometric degrees of freedom increase, making the scaling of these ideas to the inverse design of complex systems unfeasible.

An alternative approach is to utilize generative adversarial networks (GANs) \cite{goodfellow2014generative}, which have been proposed as a tool for freeform device optimization \cite{liu2018generative, jiang2019free, so2019designing}.  GANs have been of great interest in recent years and have a broad range applications, including image generation \cite{brock2018large, karras2017progressive}, image synthesis \cite{zhu2018visual}, image translation \cite{zhu2017unpaired}, and super resolution imaging \cite{ledig2017photo}. In the context of photonics inverse design, GANs are provided images of high performance devices, and after training, they can generate high performance device patterns with geometric features mimicking the training set \cite{jiang2019free}.  With this approach, devices from a trained GAN can be produced with low computational cost, but a computationally expensive training set is required.  New data-driven concepts that can reduce or even eliminate the need for expensive training data would dramatically expand the scope and practicality of machine learning-enabled device optimization.

\section{Problem Setup}
\begin{figure}[h]
  \centering
  \includegraphics[width=\linewidth]{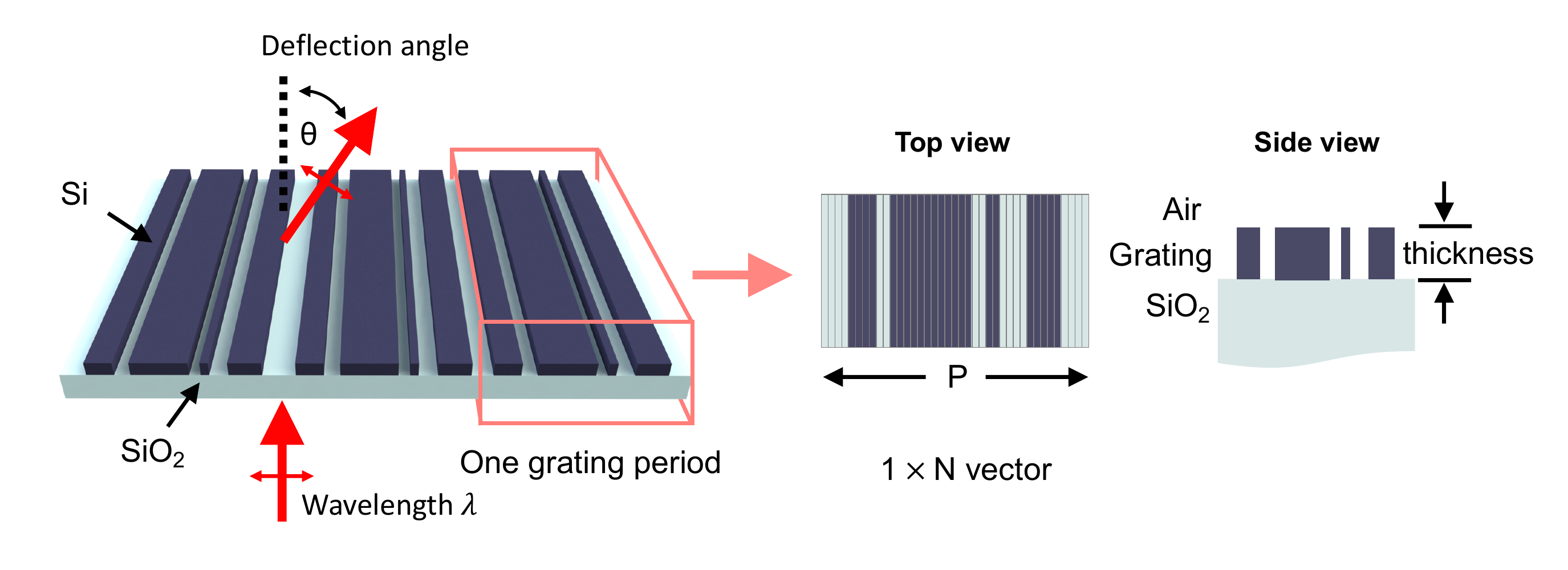}
  \caption{Schematic of a silicon metagrating that deflects normally incident TM-polarized light of wavelength $\lambda$ to an outgoing angle $\theta$. The optimization objective is to search for the metagrating pattern that maximizes deflection efficiency.}
\end{figure}
The metagratings consist of silicon nanoridges and deflect normally-incident light to the +1 diffraction order (Figure 1).  The thickness of the gratings is fixed to be 325 nm and the incident light is TM-polarized. The refractive index of silicon is taken from Ref. \cite{Silicon2008} and only the real part of the index is used to simplify the design problem.  For each period, the metagrating is subdivided into $N = 256$ segments, each possessing a refractive index value between silicon and air during the optimization process.  These refractive index values are the design variable in our problem and are specified as $\mathbf{x}$ (a $1 \times N$ vector).
Deflection efficiency is defined as the intensity of light deflected to the desired direction, defined by angle $\theta$, normalized to the incident light intensity. The deflection efficiency is a nonlinear function of index profile $\mbox{Eff} = \mbox{Eff}(\mathbf{x})$ and is governed by Maxwell's equations.  This quantity, together with the electric field profiles within a device, can be accurately solved using electromagnetic solvers. 

Our optimization objective is to maximize the deflection efficiency of the metagrating at a specific operating wavelength $\lambda$ and outgoing angle $\theta$:
\begin{equation}
    \mathbf{x}^{*}:= \argmax_{\mathbf{x} \in \{-1, 1\}^N} \ \mbox{Eff}(\mathbf{x})
    \label{Eq1}
\end{equation}
The term $\mathbf{x}^{*}$ represents the globally optimized device pattern, and it has an efficiency of $\mbox{Eff}_{max}$. We are interested in physical devices that possess binary index values in the vector: $\mathbf{x} \in \{-1, 1\}^N$, where -1 represents air and +1 represents silicon. 

\section{Methods}

\begin{figure}[h]
  \centering
  \includegraphics[width=\linewidth]{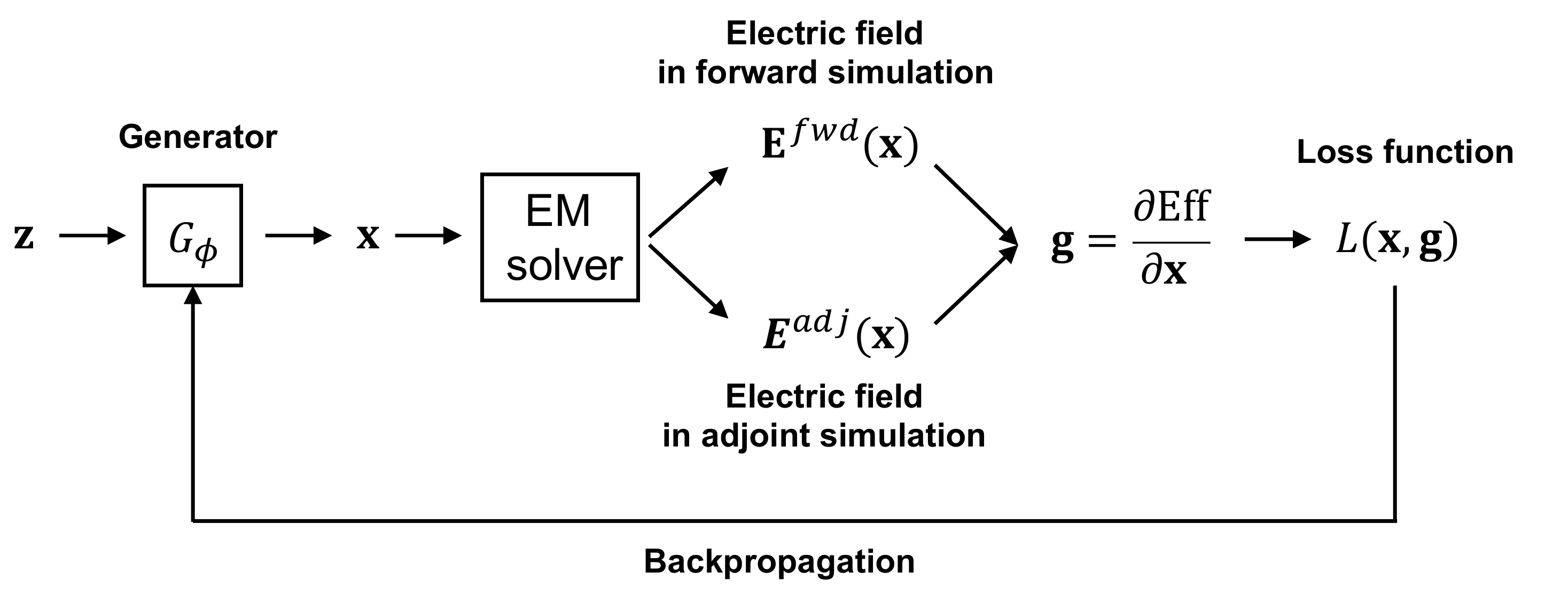}
  \caption{Schematic of generative neural network-based optimization.}
\end{figure}

Our proposed inverse design scheme is shown in Figure 2 and involves the training of a generative neural network to optimize a population of devices.  Uniquely, our scheme does not require any training set. The input of the generator is a random noise vector $\mathbf{z} \in \mathcal{U}^N(-1, 1) $ and it has the same dimension as the output device index profile $\mathbf{x} \in [-1, 1]^N$. The generator is parameterized by $\phi$, which relates $\mathbf{z}$ to $\mathbf{x}$ through a nonlinear mapping: $\mathbf{x} = G_{\phi}(\mathbf{z})$. In other words, the generator maps a uniform distribution of noise vectors to a device distribution $G_\phi : \mathcal{U}^N(-1, 1) \mapsto P_\phi$, where $P_{\phi}(\mathbf{x})$ defines the probability of generating $\mathbf{x}$ in the device space $\mathcal{S} = [-1, 1]^N$. We frame the objective of the optimization as maximizing the probability of generating the globally optimized device in $\mathcal{S}$: 
\begin{equation}
    \mathbf{\phi}^{*}:= \argmax_{\mathbf{\phi}} \ \int_{\mathcal{S}} \delta \left( \mbox{Eff}(\mathbf{x}) - \mbox{Eff}_{max}\right) \cdot P_\phi(\mathbf{x}) d\mathbf{x}
    \label{Eq2}
\end{equation}

\subsection{Loss Function Formulation}
While our objective function above is rigorous, it cannot be directly used for network training due to two reasons.  The first is that the derivative of the $\delta$ function is nearly always zero. To circumvent this issue, we express the $\delta$ function as the following: 
\begin{equation}
    \delta \left( \mbox{Eff}(\mathbf{x}) - \mbox{Eff}_{max}\right) = \lim_{\sigma \rightarrow 0}\frac{1}{\sqrt{\pi}\sigma}\exp{\left[-\left(\frac{\mbox{Eff}(\mathbf{x}) - \mbox{Eff}_{max}}{\sigma}\right)^2 \right]}
    \label{Eq3}
\end{equation}
By substituting the $\delta$ function with this Gaussian form and leaving $\sigma$ as a tunable parameter, we relax Equation \ref{Eq2} and it becomes:  
\begin{equation}
    \mathbf{\phi}^{*}:= \argmax_{\mathbf{\phi}} \ \int_{\mathcal{S}} \exp{\left[-\left(\frac{\mbox{Eff}(\mathbf{x}) - \mbox{Eff}_{max}}{\sigma}\right)^2 \right]} \cdot P_\phi(\mathbf{x}) d\mathbf{x}
    \label{Eq4}
\end{equation}
As we will see later, the inclusion of $\sigma$ as a tunable hyperparameter turns out to be important for stabilizing the network training process in the limit of training with a finite batch size.

The second reason is that the objective function depends on $\mbox{Eff}_{max}$, which is unknown.  To address this problem, we approximate Equation \ref{Eq4} with a different function, namely the exponential function:
\begin{align}
    \mathbf{\phi}^{*}:&= \argmax_{\mathbf{\phi}} \ \int_{\mathcal{S}} \exp{\left(\frac{\mbox{Eff}(\mathbf{x}) - \mbox{Eff}_{max}}{\sigma}\right)} \cdot P_\phi(\mathbf{x}) d\mathbf{x}
    \label{Eq5a}
\end{align} 
This approximation is valid because $P_\phi(\mathbf{x} \ |\  \mbox{Eff}(\mathbf{x}) > \mbox{Eff}_{max}) = 0$ and our new function only needs to approximate that in Equation \ref{Eq4} for efficiency values less than $\mbox{Eff}_{max}$.  With this approximation, we can remove $\exp{\left(-\mbox{Eff}_{max}/\sigma\right)}$  from the integral:
\begin{align}    
    \mathbf{\phi}^{*}:&= \argmax_{\mathbf{\phi}} \ A\int_{\mathcal{S}} \exp{\left(\frac{\mbox{Eff}(\mathbf{x})}{\sigma}\right)} \cdot P_\phi(\mathbf{x}) d\mathbf{x}
    \label{Eq5b}
\end{align}
$A = \exp(-\mbox{Eff}_{max}/\sigma)$ now becomes a normalization constant and does not require explicit evaluation. Alternatives to the exponential function can be considered and tailored depending on the specific optimization problem. For this study, we will use Equation \ref{Eq5b}. 

In practice, it is not possible to evaluate Equation \ref{Eq5b} over the entire design space $\mathcal{S}$.  We instead sample a batch of devices $\{\mathbf{x}^{(m)}\}_{m=1}^{M}$ from $P_\phi$, which leads to further approximation of the objective function:
\begin{align}
    \mathbf{\phi}^{*}:&= \argmax_{\mathbf{\phi}} \underset{\mathbf{x} \sim P_\phi}{\mathbb{E}} \exp{\left(\frac{\mbox{Eff}(\mathbf{x})}{\sigma}\right)} \\
    &\approx \argmax_{\mathbf{\phi}} \frac{1}{M}\sum_{m=1}^{M}\exp{\left(\frac{\mbox{Eff}(\mathbf{x}^{(m)})}{\sigma}\right)}
\end{align}   
We note that the deflection efficiency of device $\mathbf{x}$ is calculated using an electromagnetic solver, such that $\mbox{Eff}(\mathbf{x})$ is not directly differentiable for backpropagation. To bypass this problem, we use the adjoint variables method to compute the efficiency gradient with respect to the refractive indices for device $\mathbf{x}$: $\mathbf{g} = \frac{\partial \mbox{Eff}}{\partial \mathbf{x}}$ (Figure 2). Details pertaining to these gradient calculations can be found in other inverse design papers \cite{Vuckovic2015, SFan2018, sell2017large}.  To summarize, electric field profiles within the device layer are calculated using two different electromagnetic  excitation conditions.  The first is the forward simulation, in which $\mathbf{E}^{fwd}$ are calculated by propagating a normally-incident electromagnetic wave from the substrate to the device, as shown in Figure 1.  The second is the adjoint simulation, in which $\mathbf{E}^{adj}$ are calculated by propagating an electromagnetic wave in the direction opposite of the desired outgoing direction. The efficiency gradient $\mathbf{g}$ is calculated by integrating the overlap of those electric field terms:
\begin{align}
    \mathbf{g} = \frac{\partial \mbox{Eff}(\mathbf{x})}{\partial \mathbf{x}} \propto \mbox{Re}(\mathbf{E}^{fwd}\cdot \mathbf{E}^{adj})
\end{align}
Finally, we use our adjoint gradients and objective function to define the loss function $L = L(\mathbf{x}, \mathbf{g}, \mbox{Eff})$.  Our goal is to define $L$ such that minimizing $L$ is equivalent to maximizing the objective function $\frac{1}{M}\sum_{m=1}^{M}\exp{\left(\frac{\mbox{Eff}(\mathbf{x}^{(m)}) }{\sigma}\right)}$ during generator training.  With this definition, $L$ must satisfy $-\frac{\partial L}{\partial \mathbf{x^{(m)}}} = \frac{1}{M}\frac{\partial }{\partial \mathbf{x}^{(m)}}\exp{\left(\frac{\mbox{Eff}(\mathbf{x}^{(m)})}{\sigma}\right)}$ and is defined as:
\begin{equation}
    L(\mathbf{x}, \mathbf{g}, \mbox{Eff}) = -\frac{1}{M}\sum_{m=1}^{M}  \frac{1}{\sigma} \exp{\left(\frac{\mbox{Eff}^{(m)}}{\sigma}\right)}\ \mathbf{x}^{(m)}\cdot \mathbf{g}^{(m)}
    \label{Eq9}
\end{equation}
$\mbox{Eff}^{(m)}$ and $\mathbf{g}^{(m)}$ are treated as independent variables calculated from electromagnetic simulations and have no dependence on $\mathbf{x}^{(m)}$. Finally, we add a regularization term $-|\mathbf{x}|\cdot(2 - |\mathbf{x}|)$ to $L$ to ensure that the generated patterns are binary. This term reaches a minimum when the generated patterns are fully binarized. A coefficient $\gamma$ is introduced to balance binarization with efficiency enhancement, and we have as our final loss function:
\begin{equation}
    L(\mathbf{x}, \mathbf{g}, \mbox{Eff}) = -\frac{1}{M}\sum_{m=1}^{M} \frac{1}{\sigma} \exp{\left(\frac{\mbox{Eff}^{(m)}}{\sigma}\right)}\ \mathbf{x}^{(m)}\cdot \mathbf{g}^{(m)}   -\gamma\cdot\frac{1}{M}\sum_{m=1}^{M} |\mathbf{x}^{(m)}|\cdot(2 - |\mathbf{x}^{(m)}|)
    \label{Eq10}
\end{equation}

\subsection{Network Architecture}
The architecture of the generative neural network is adapted from DCGAN \cite{radford2015unsupervised}, which comprises 2 fully connected layers, 4 transposed convolution layers, and a Gaussian filter at the end to eliminate small features. LeakyReLU is used for all activation functions except for the last layer, which uses a tanh activation function. We also add dropout layers and batchnorm layers to enhance the diversity of the generated patterns. Periodic paddings are used to account for the fact that the devices are periodic structures. 
    
\subsection{Training Procedure}

\begin{algorithm}[H]
\SetAlgoLined
\SetKwInOut{Parameter}{Parameters}
\Parameter{$M$, batch size. $\sigma$, loss function coefficient. $\alpha$, learning rate. $\beta_1$ and $\beta_2$, momentum coefficients used in Adam. $\gamma$, binarization coefficient.}
 initialization\;
 \While{i < Total iterations}{
  Sample $\{\mathbf{z}^{(m)}\}_{m=1}^{M} \sim \mathcal{U}^N(-1, 1)$\;
  $\{ \mathbf{x}^{(m)} = G_{\phi}(\mathbf{z}^{(m)})\}_{m=1}^{M}$, device samples\;
  $\{\mathbf{g}^{(m)}\}_{m=1}^{M}$, $\{\mbox{Eff}^{(m)}\}_{m=1}^{M}$ $\leftarrow$ forward and adjoint simulations\;
  $g_\phi \leftarrow \nabla_{\phi}\left[ \frac{1}{M}\sum_{m=1}^{M}  \frac{1}{\sigma} \exp{\left(\frac{\mbox{Eff}^{(m)}}{\sigma}\right)}\ \mathbf{x}^{(m)}\cdot \mathbf{g}^{(m)} \;
  + \gamma\cdot\frac{1}{M}\sum_{m=1}^{M} |\mathbf{x}^{(m)}|\cdot(2 - |\mathbf{x}^{(m)}|) \right] $\;
  $\phi \leftarrow \phi + \alpha\cdot\mbox{Adam}(\phi, g_\phi)$\;
 }
 $\mathbf{x}^* \leftarrow \argmax_{ \mathbf{x} \in \{ \mathbf{x}^{(m)} | \mathbf{x}^{(m)} \sim P_{\phi^*} \}_{m=1}^{M}} \mbox{Eff}(\mathbf{x})$
 \caption{Generative neural network-based optimization}
 \label{Ag1}
\end{algorithm}

The training procedure is shown in Algorithm \ref{Ag1}. The Adaptive Moment Estimation (Adam) algorithm, which is a variation of gradient descent,  is used to optimize the network parameters $\phi$. $\beta_1$ and $\beta_2$ are two hyperparameters used in Adam \cite{kingma2014adam}.  Initially, with the use of an identity shortcut \cite{he2016deep}, the device distribution $P_{\phi}$ is approximately a uniform distribution over the whole device space $\mathcal{S}$. During the training process, $P_{\phi}$ is continuously refined and maps more prominently to high-efficiency device subspaces. When the generator is done training, the devices produced from the generator have a high probability to be highly efficient.  The final optimal device design is determined by generating a batch of devices from the fully trained generator $\{ \mathbf{x}^{(m)} | \mathbf{x}^{(m)} \sim P_{\phi^*} \}_{m=1}^{M}$, simulating each of those devices, and selecting the best one. 

\subsection{Comparison with gradient-based topology optimization}
In gradient-based topology optimization, a large number of local optimizations are used to search for the global optimum.  For each run, device patterns are randomly initialized, and a local search in the design space is performed using gradient descent. The highest efficiency device among those optimized devices is taken as the final design. With this approach, many devices get trapped in local optima or saddle points in $\mathcal{S}$, and the computational resources used to optimize those devices do not contribute to finding or refining the globally optimal device.  Additionally, finding the global optimum in a very high dimensional space can require an exceedingly large number of individual optimization runs.  GLOnets are qualitatively different, as they optimize a distribution of devices to perform global optimization. As indicated in Equation \ref{Eq10}, each device sample $\mathbf{x}^{(m)}$ is weighted by the term $\exp{(\mbox{Eff}^{(m)}/\sigma)}$, which biases generator training towards higher efficiency devices and pushes $P_{\phi}$ towards more favorable design subspaces. In this manner, computational resources are not wasted optimizing devices within subspaces possessing low-efficiency local optima. 

\section{Numerical Experiments}
\subsection{A Toy Model}
We first perform GLOnet-based optimization on a simple test case, where the input $\mathbf{z}$ and output $\mathbf{x}$ are two dimensional. The "efficiency" function $\mbox{Eff}(\mathbf{x})$ is defined as:
\begin{equation}
    \mbox{Eff}(x_1,x_2) = \exp(-2x_1^2)\cos{(9x_1)} + \exp(-2x_2^2)\cos{(9x_2)}
\end{equation}
This function is non-convex and has many local optima and one global optimum at (0, 0). We use Algorithm 1 to search for the global optimum. Hyperparameters are chosen to be $\alpha = 1e-3, \beta_1 = 0.9, \beta_2=0.999, a = 30,$ and $\sigma = 0.5$, and the batch size $M =100$ is fixed throughout network training. The generator is trained for 150 iterations and the generated samples over the course of training are shown as red dots in Figure 3. Initially, the generated "device" distribution is spread out over the $\mathbf{x}$ space, and it then gradually converges to a cluster located at the global optimum.  In the training run shown, no device is trapped in any local optima. Upon training 100 distinct GLOnets, 96 of them successfully produced the globally optimized device.

\begin{figure}[h]
  \centering
  \includegraphics[width=\linewidth]{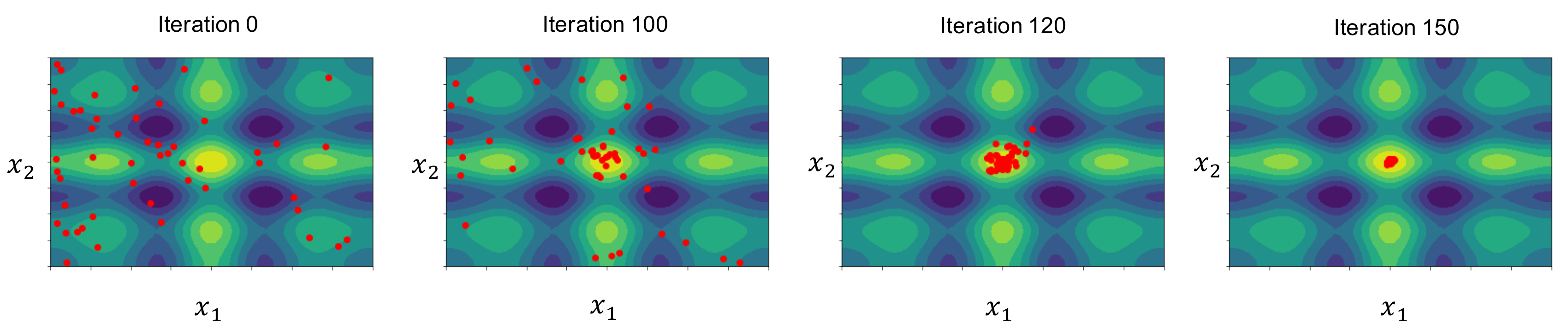}
  \caption{Results from a toy model test. Samples generated from the generator, shown as red dots, evolve in the $[-1, 1]^2$ space over the course of training.}
\end{figure}

\subsection{Inverse design of metagratings}
We next apply our algorithm to the inverse design of 63 different types of metagratings, each with differing operating wavelengths and deflection angles.  The wavelengths $\lambda$ range from 800 nm to 1200 nm, in increments of 50 nm, and the deflection angles $\theta$ range from 40 degrees to 70 degrees, in increments of 5 degrees.  Unlike our conditional GLOnet in Ref. \cite{jiang2019global}, where many different types of metagratings are simultaneously designed using a single network, we use distinct unconditional GLOnets to design each device type operating for specific wavelength and deflection angle parameters.

\begin{figure}[ht]
  \centering
  \includegraphics[width=\linewidth]{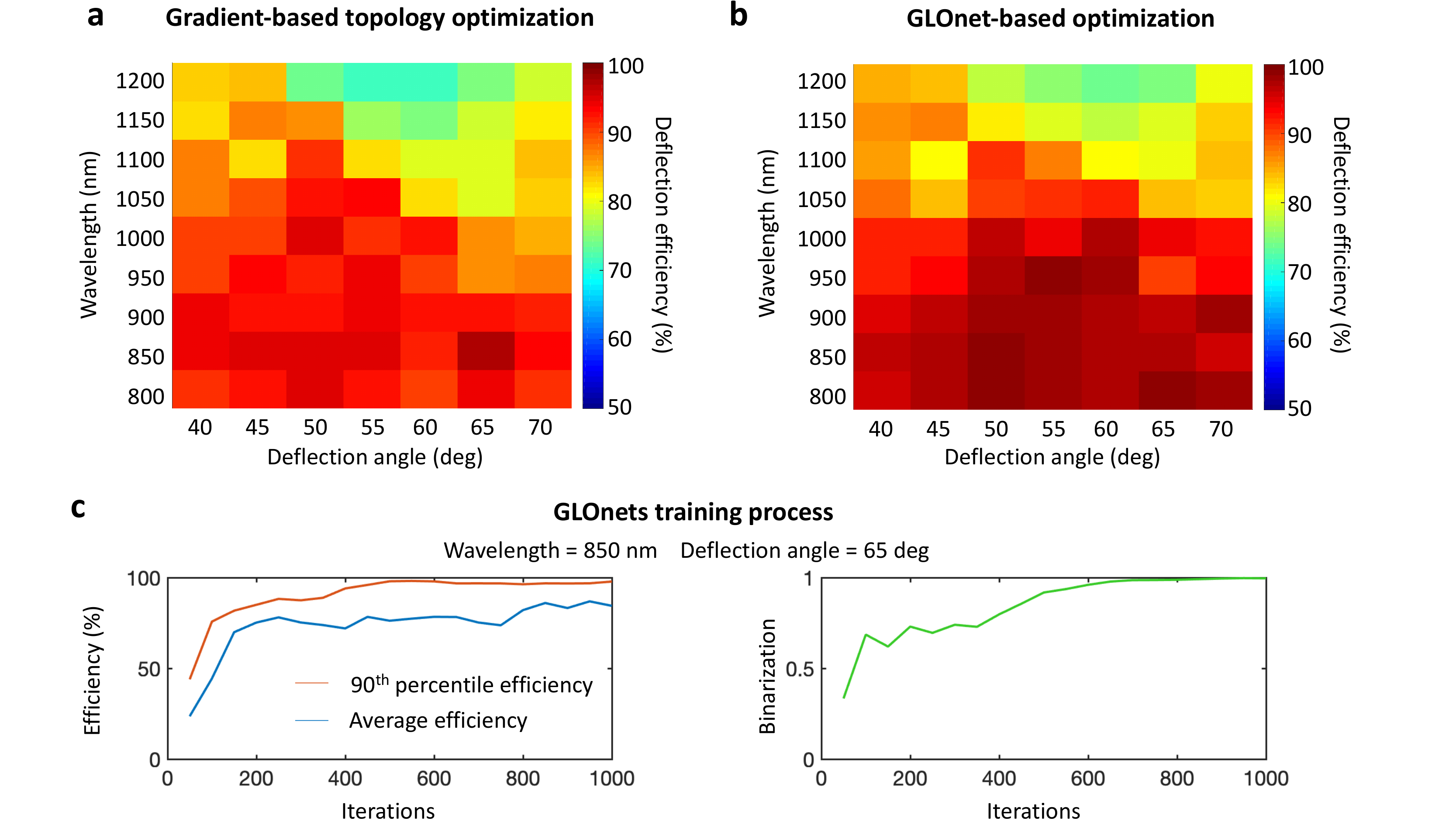}
  \caption{(a) Plot of efficiency for devices operating with different wavelength and angle values, designed using gradient-based topology optimization. For each wavelength and angle combination, 500 individual topology optimizations are performed and the highest efficiency device is used for the plot.  (b) Plot of efficiency for devices designed using GLOnet-based optimization. For each wavelength and angle combination, 500 devices are generated and the highest efficiency device is used for the plot. (c) Training process of GLOnets. The figure on the left shows the 90th percentile efficiency and average efficiency of the device batch over the course of training. The figure on the right shows the binarization degree of generated devices, which is defined as $\sum_{i=1}^{N}|\mathbf{x}_i|/N$.}
  \label{Fig4}
\end{figure}

\begin{figure}[ht]
  \centering
  \includegraphics[width=\linewidth]{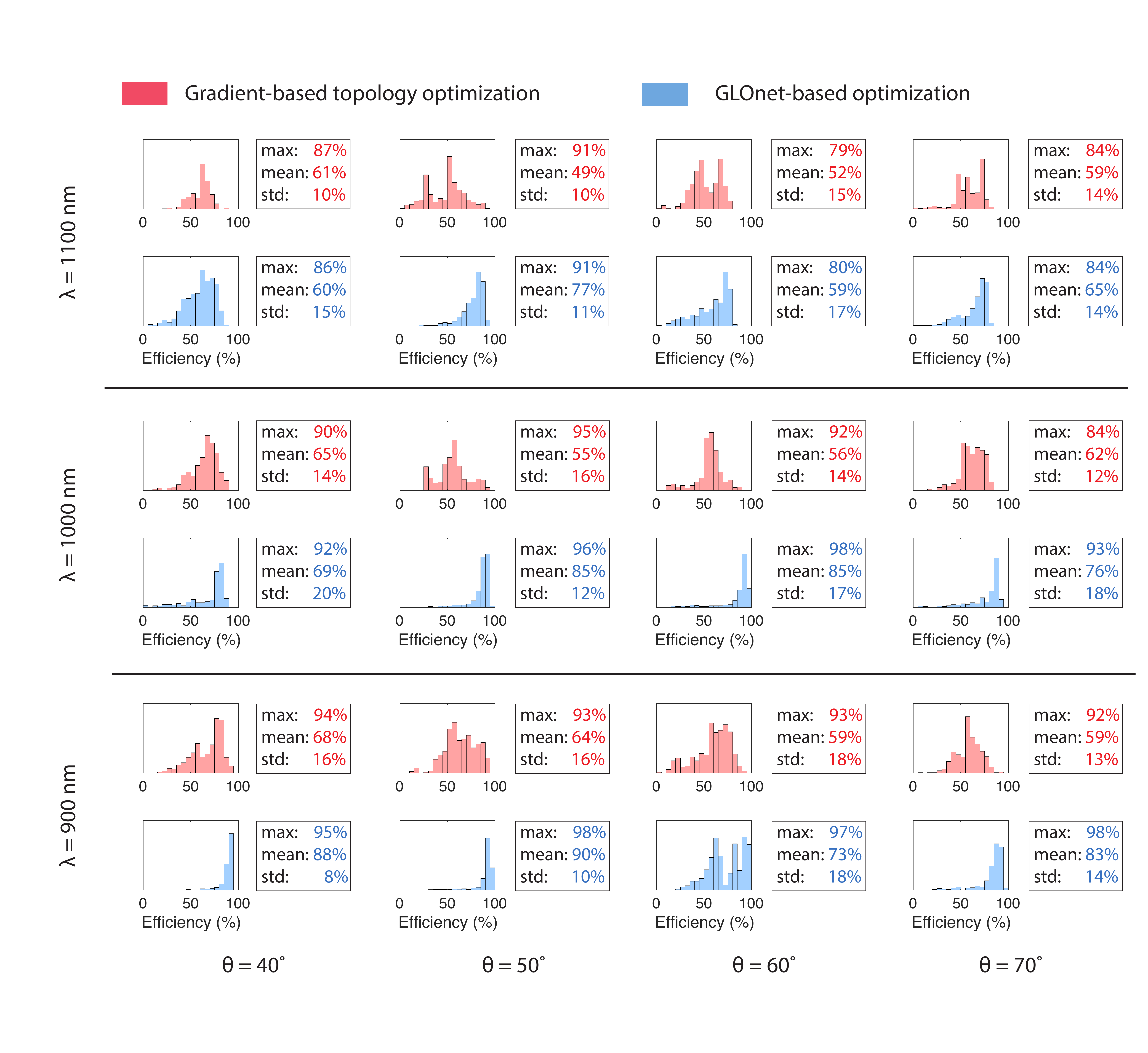}
  \caption{Efficiency histograms of 500 devices designed using gradient-based topology optimization (red) and GLOnet-based optimization (blue).  The statistics of device efficiencies in each histogram are also displayed.  For most wavelength and angle values, the efficiency distributions from GLOnets are narrower and have higher maximum values compared to those from gradient-based topology optimization.}
   \label{Fig5}
\end{figure}

\subsubsection{Implementation details} 
The hyperparameters we use are $\alpha = 0.05, \beta_1 = 0.9, \beta_2=0.99, \sigma = 0.5,$ and $\gamma = 0.2$. The batch size is 100. To prevent vanishing gradients when the generated patterns are binarized as $\mathbf{x} \in \{-1, 1\}^N$, we specify the last activation function to be $1.05*\tanh$.

For each combination of wavelength and angle, we train the generator for 1000 iterations. Upon completion of network training, 500 different values of $\mathbf{z}$ are used to generate 500 different devices.  All devices are simulated and the highest efficiency device is taken as the final design.

The network is implemented using the pytorch-1.0.0 package. The forward and adjoint simulations are performed using the Reticolo RCWA \cite{hugonin2005reticolo} electromagnetic solver in MATLAB. The network is trained on an Nvidia Titan V GPU and 4 CPUs, and it takes 10 minutes for one device optimization. Our code implementation can be found at: \href{https://github.com/jiaqi-jiang/GLOnet.git}{https://github.com/jiaqi-jiang/GLOnet.git}.

\subsubsection{Baseline} 
We benchmark our method with gradient-based topology optimization. For each design target $(\lambda, \theta)$, we start with 500 random gray-scale vectors and iteratively optimize each device using efficiency gradients calculated from forward and adjoint simulations. A threshold filter binarizes the device patterns. Each initial dielectric distribution is optimized for 200 iterations, and the highest efficiency device among 500 candidates is taken as the final design. The computational budget is set to be the same used for GLOnets training to facilitate a fair comparison.

\subsubsection{Results}
The efficiencies of the best devices designed using gradient-based topology optimization and GLOnets are shown in Figure \ref{Fig4}. 90\% of the best devices from GLOnets have higher or the same efficiencies compared to the best devices produced from gradient-based topology optimization. 98\% of the best devices from GLOnets have efficiencies within 5\% of the best devices from gradient-based topology optimization. For wavelengths and angles for which GLOnets perform worse than gradient-based topology optimization, we can perform multiple network trainings or further tune the batch size and $sigma$ to get better GLOnet results. The efficiency histograms from GLOnets and gradient-based topology optimization, for select wavelength and angle pairs, are displayed in Figure \ref{Fig5}. For most cases, efficiency histograms produced from our method have higher average efficiencies and maximal efficiencies, indicating that low-efficiency local optima are often avoided during the training of the generator. Device patterns generated by a well-trained generator are shown in Figure \ref{Fig4}d and all look relatively similar, indicating that the generator has collapsed around a particular high-efficiency device topology irrespective of the input values of $\mathbf{z}$.

\subsubsection{GLOnet Stability}
\begin{figure}[H]
  \centering
  \includegraphics[width=\linewidth]{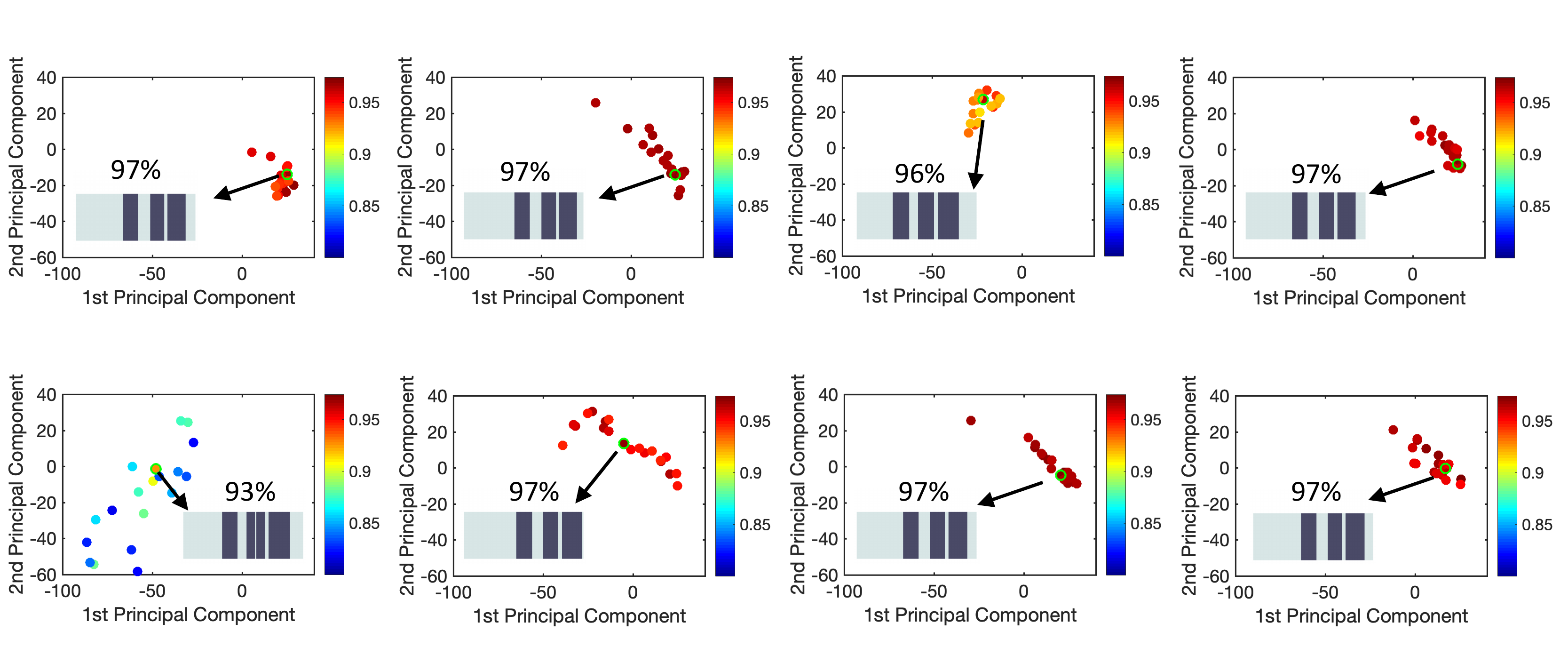}
  \caption{PCA visualization of GLOnet-optimized devices. Eight unconditional GLOnets are trained independently and the top 20 devices of each GLOnet are visualized.  The pattern and efficiency of the best device in each plot are shown as insets.}
   \label{Fig6}
\end{figure}

To validate the stability of GLOnet-based optimization, we train eight unconditional GLOnets independently for the same wavelength ($\lambda = 850$ nm) and deflection angle ($\theta = 65$ degrees). For each trained GLOnet, we generate 500 devices and visualize the top 20 devices in a 2D plane using principle component analysis (PCA) (Figure \ref{Fig6}). The principle basis is the same for all eight figures and is calculated using the top 20 devices of each GLOnet for a total of 160 devices.  Six of the eight GLOnets converge to the same optimum, which is a device with 97\% efficiency, while one GLOnet converges to a nearby optimum, which is a device with 96\% efficiency.  While we cannot prove that the device with 97\% efficiency is globally optimal, the consistent convergence of GLOnet to this single optimum is suggestive that the network is finding the global optimum.  At the very least, this demonstration shows that GLOnets have the potential to consistently generate exceptionally high performance devices.

\subsubsection{Discussion of hyperparameter $\sigma$ and batch size}
In principle, $\sigma$ approaching zero could be used if the entire design space could be sampled to train the neural network.  In this case, the globally optimized structure would be sampled and be the only device that contributes to neural network training, pushing the response of the network towards our desired objective response.  However, the design space is immense and infeasible to probe in its entirety.  Furthermore, this scenario would lead to the direct readout of the globally optimized device, negating the need to perform an optimization.  

In practice, we can only realistically process small batches of devices that comprise a very small fraction of the total design space during network training.  For many of these iterations, the best device within each batch will only be in locally optimal regions of the design space.  To prevent the network from getting trapped within these local optima, we specify $\sigma$ to be finite, which adds noise to the training process.  In our loss function, this noise manifests in the form $\exp(-\text{Eff}/\sigma)$.  This exponential expression has a Boltzmann form and $\sigma$ can therefore be treated as an effective temperature.  In a manner analogous to the process of simulated annealing, $\sigma$ can be modulated throughout the training process.

\begin{figure}[H]
  \centering
  \includegraphics[width=\linewidth]{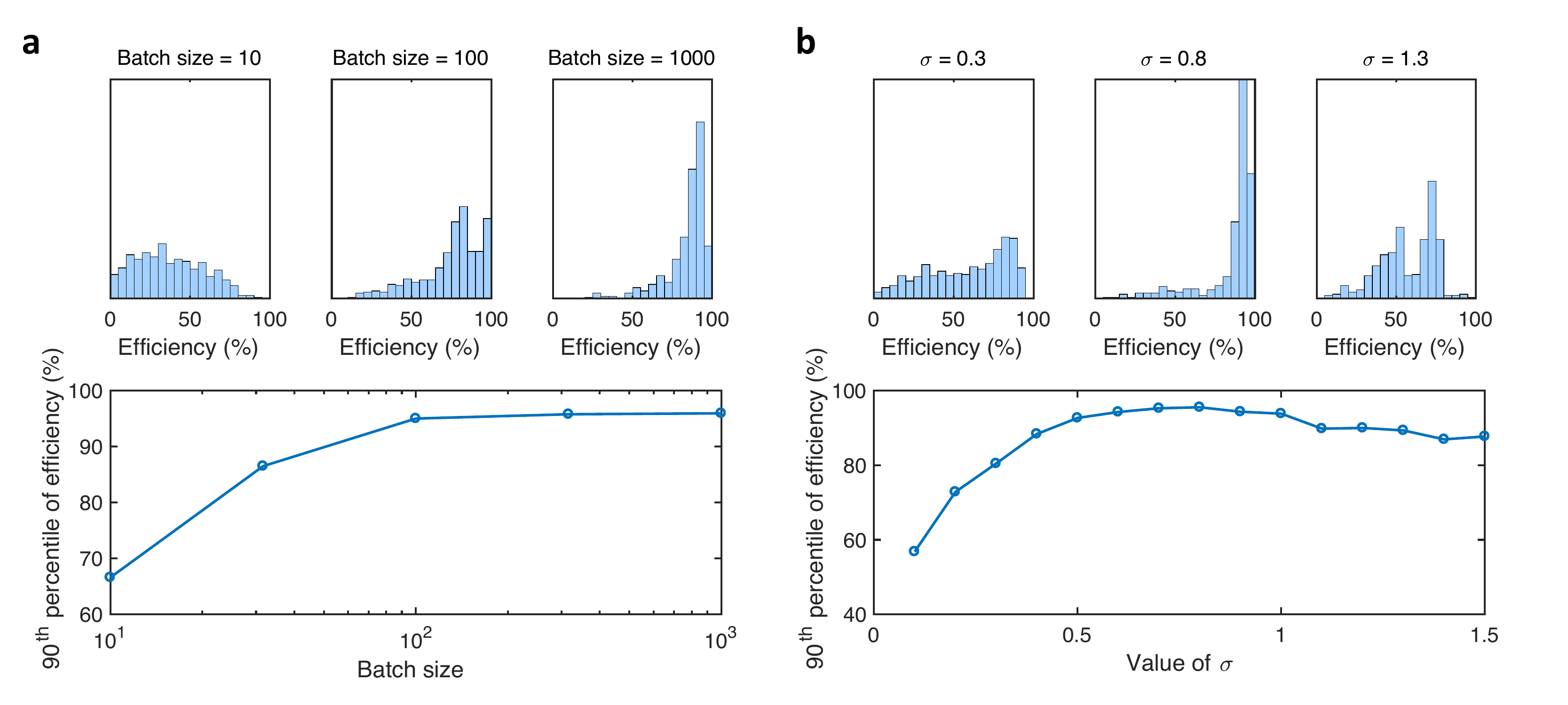}
  \caption{Performance of the unconditional GLOnet for different values of (a) batch size and (b) $\sigma$.}
   \label{Fig7}
\end{figure}

The impact of batch size and $\sigma$ on GLOnet performance for $\lambda = 850$ nm and $\theta = 65$ degrees is summarized in Figure \ref{Fig7}.  In Figure \ref{Fig7}a, $\sigma$ is fixed to be 0.5 and the batch size is varied from 10 to 1000 devices per iteration.  When the batch size is too small, the design space is undersampled, which increases the difficulty of finding the global optimum.  As the batch size is increased, the performance of the GLOnet starts to saturate such that the design space is oversampled, leading to a waste of computational resources.  For this particular GLOnet, a proper batch size that balances optimization capability with resource management is 100.  

Figure \ref{Fig7}b summarizes the impact of $\sigma$ on GLOnet training, given a fixed batch size of 100 devices.  The plot indicates that a proper range of $\sigma$ that produces the highest efficiency devices is between 0.5 to 1.0.  When $\sigma$ is less than 0.5, there is insufficient noise in the training process and the network gets more easily trapped within local optima, particularly early in the training process.  When $\sigma$ becomes larger than 1, the performance of the GLOnet begins to deteriorate because low efficiency devices contribute more significantly in the training process, leading to excess noise. 

The optimal batch size and $\sigma$ values are highly problem dependent and require tuning for each optimization objective.  For example, proper GLOnet optimization within a design space with relatively few local optima can be achieved with relatively small batch sizes and small values of $\sigma$.  The proper selection of these hyperparameters is not intuitive and requires experience and parametric sweeps.

\section{Comparison with evolution strategies (ES)}
Evolutionary strategies represent classical global optimization strategies.  One such algorithm is the genetic algorithm, which have been applied to many types of photonic design problems, including metasurface design \cite{egorov2017genetically}.  Compared to our approach, genetic algorithms are not efficient and require many thousands of iterations to search for even a simple optimal device structure. The difficulty is due to the complicated relationship between optical response and device geometry, governed by Maxwell's equations. Methods like ours, which incorporate gradients, can more efficiently locate favorable regions of the design space because gradients provide clear, non-heuristic instruction on how to improve device performance. 

Another ES algorithm is the Covariance Matrix Adaptation Evolution Strategy (CMA-ES), which is a probability distribution-based ES algorithm. CMA-ES assumes an explicit form of the probability distribution of the design variables (e.g. multivariate normal distribution), which is typically parameterized by several terms. Our algorithm has two main differences compared with CMA-ES. First, instead of defining an explicit probability distribution, we define an explicit generative model parameterized by the network parameters. The probability distribution in our algorithm is therefore implicit and has no assumed form. This is important as there is no reason why the probability distributions of the design variables should have a simple, explicitly defined form such as the multivariate normal distribution.  Second, CMA-ES is derivative-free, but our algorithm uses gradients and is therefore more efficient at generating device populations in the desirable parts of the design space.

\section{Conclusions and Future Directions}
In this paper, we present a generative neural network-based global optimization algorithm for metasurface design. Instead of optimizing many devices individually, which is the case for gradient-based topology optimization, we reframe the global optimization problem as the training of a generator. The efficiency gradients of all devices generated each training iteration are used to collectively improve the performance of the generator and map the noise input to favorable regions of the device subspace.

An open topic of future study is understanding how to properly select and tune the network hyperparameters dynamically during network training.  We anticipate that, as the distribution of generated devices converges to a narrow range of geometries over the course of network training, the batch size can be dynamically decreased, leading to computational savings.  We also hypothesize that dynamically decreasing $\sigma$ can help further stabilize the GLOnet training process.  These variations in batch size and $\sigma$ can be predetermined prior to network training or be dynamically modified using feedback during the training process.

We are also interested in applying our algorithm to more complex systems, such as 2D or 3D metasurfaces, multi-function metasurfaces, and other photonics design problems. A deeper understanding of loss function engineering will be necessary for multi-function metasurfaces design, which requires optimizing multiple objectives simultaneously. We envision that our algorithm has strong potential to solve inverse design problems in other domains of the physical sciences, such as mechanics and electronics.  

\section*{Acknowledgement}
 The simulations were performed in the Sherlock computing cluster at Stanford University. This work was supported by the U.S. Air Force under Award Number FA9550-18-1-0070, the Office of Naval Research under Award Number N00014-16-1-2630, and the David and Lucile Packard Foundation.

\end{document}